%% file: paper-PRB.tex
\documentclass[aps,prb, twocolumn, showpacs, showkeys,unsortedaddress]{revtex4-1}
\usepackage{color}
\usepackage[utf8]{inputenc}
\usepackage[dvips]{graphicx}
\usepackage{wrapfig}
\usepackage{amsmath}
\usepackage{amsfonts}
\usepackage{amssymb}
\usepackage{subfigure}
\usepackage{slashbox}
\usepackage{vector}

\newcommand{\bra}[1]{\langle #1|}
\newcommand{\ket}[1]{|#1\rangle}

\definecolor{grey}{rgb}{0.8,0.8,0.8}
\definecolor{note}{rgb}{0.9,.1,.1} % notes Sebastian
\definecolor{sebnote}{rgb}{.3,.2,.9} % notes Sebastian

\begin{document}
\title{Coulomb interaction effects and electron spin relaxation in the 1d Kondo lattice model}
\affiliation{Physics Department, Arnold Sommerfeld Center for Theoretical Physics and Center for 
NanoScience, Ludwig-Maximilians-Universit\"at M\"unchen, 80333 M\"unchen, Germany}
\affiliation{Institut f\"ur Theorie der Statistischen Physik and
JARA-Fundamentals of Future Information Technology, RWTH Aachen University, D-52056 Aachen, Germany}
\affiliation{School of Physical Sciences, University of Queensland, QLD 4072, Australia}

\author{Sebastian \surname{Smerat}}
\email[]{Sebastian.Smerat@physik.uni-muenchen.de}
\affiliation{Physics Department, Arnold~Sommerfeld~Center~for~Theoretical~Physics, Ludwig-Maximilians-Universit\"at M\"unchen, 80333 M\"unchen, Germany}
\author{Herbert Schoeller}
\affiliation{Institut f\"ur Theorie der Statistischen Physik and
JARA-Fundamentals of Future Information Technology, RWTH Aachen University, D-52056 Aachen, Germany}
\author{Ian~P.~McCulloch}
\affiliation{School of Physical Sciences, University of Queensland, QLD 4072, Australia}
\author{Ulrich Schollw\"ock}
\affiliation{Physics Department, Arnold Sommerfeld Center for Theoretical Physics and Center for 
NanoScience, Ludwig-Maximilians-Universit\"at M\"unchen, 80333 M\"unchen, Germany}

\date{\today}
\begin{abstract}
We study the effects of the Coulomb interaction in the one dimensional
Kondo lattice model on the phase diagram, the static magnetic
susceptibility and electron spin relaxation. We show that
onsite Coulomb interaction supports ferromagnetic order and nearest
neighbor Coulomb interaction drives, depending on the electron filling,
either a paramagnetic or ferromagnetic order. Furthermore we calculate
electron quasiparticle life times, which can be related to electron
spin relaxation and decoherence times, and explain their dependence 
on the strength of interactions and the electron filling in order to
find the sweet spot of parameters where the relaxation time is maximized.
We find that effective exchange processes between the electrons dominate
the spin relaxation and decoherence rate.
\end{abstract}
\maketitle
%=====================================================
%
%                   INTRODUCTION
%
%=====================================================
\section{Introduction}
Recently, the interest in nanoscale systems has been rapidly
increasing. Among them are ${}^{13}C$ carbon nanotubes,
\cite{Marcus09, Braunecker09}  nanowires \cite{Reininghaus06, Rodrigues03}
and carbon nanotubes filled with endohedral fullerenes or molecular
magnets \cite{Krive06}. The above mentioned systems have in common,
that they consist of local spins (electron or nuclear spins)
which interact via exchange interaction with itinerant conduction
electrons. These are exactly the constituents of the one dimensional 
Kondo lattice model \cite{Schrieffer66,Tsunetsugu97} (\emph{KLM}). 
To make these materials available for spin electronics or quantum 
information processing it is necessary to understand their properties 
in detail: ground state (e.g. magnetic order), spectral 
(e.g. dispersion relation of electrons) and dynamical 
(e.g. non-equilibrium, spin relaxation/decoherence) properties.

\begin{figure}[t]
	\includegraphics[width=4.3cm]{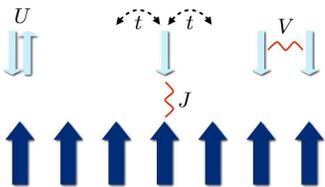}	
	\caption{\emph{(Color online)} The Kondo lattice model. The conduction electrons are 
          depicted in the upper row (red) and the localized electrons are depicted as bold arrows 
          in the lower row (grey).}
	\label{fig:klm}
\end{figure}

Interaction between the local spins in the KLM is generated
effectively due to the hopping $t$ of electrons and an onsite direct 
spin exchange $J$ between the itinerant and localized spins, see Fig.~\ref{fig:klm}.
This interaction is a result of the competition of onsite singlet 
formation and an effective RKKY (Ruderman-Kittel-Kasuya-Yosida) 
interaction \cite{Ruderman1954}. The order of the local spins due 
to the interaction is captured in the phase diagram of the KLM, 
\cite{Tsunetsugu93, Tsunetsugu97, Honner97, McCulloch01, McCulloch02} 
which is basically divided into three phases depending on $J/t$ and 
the electron filling $n$ ($n=1$ is half filling). At $n=1$ the system 
turns out to order anti-ferromagnetically for arbitrary coupling strength. 
A ferromagnetic (FM) phase is established, if either $J$ is large enough or $n$ is small 
enough. \cite{Sigrist91} Otherwise the local spin lattice is in the
paramagnetic (PM) phase, because then the effective RKKY interaction dominates the system.

The mechanism of ferromagnetism in the KLM can also be understood in terms of 
an electron quasiparticle picture, where the quasiparticle is the so called 
\emph{spinpolaron}\cite{Richmond70, Shastry81}, see Fig.~\ref{fig:qubit}a. For 
a given FM order of the local spins in a 1d system it was shown that the 
itinerant electrons and the magnons of the local spin bath form a bound 
spinpolaron state which is detectable in transport measurements and was 
proposed as a long-living correlated many-body spin state \cite{Reininghaus06} 
forming possibly one part of a many-body spin qu-bit. In Ref.~\onlinecite{Sigrist91} 
it was shown for the case of a single conduction electron that a spinpolaron develops with  a huge extent over the whole lattice leading to FM order in the ground state. 
In Ref.~\onlinecite{Smerat09} this was extended to finite electron fillings and it was 
shown that long quasiparticle life times are connected with FM order of the 
local spins. In Ref.~\onlinecite{Trebst06}, the quasiparticle dynamics of the half 
filled KLM (n=1) have been examined as well. By means of a strong coupling expansion 
up to 11th order it has been possible to calculate the quasiparticle dispersion relation 
to good accuracy and it could be shown that the quasiparticles behave like nearly localized 
f-electrons due to the strong correlation of the conduction and localized electrons. 

It is known that the main relaxation and decoherence source of single electron spins 
in semiconductor based quantum dots arises from interactions with the nuclear spin 
background. \cite{Loss98,Hanson07,Fischer09}
An appropriate path to diminish the relaxation is the application of a large magnetic field, whereas
the decoherence rate is reduced by state distribution narrowing. \cite{Coish04,Coish08}
However, the initial preparation of the nuclear bath in a pure state (e.g. full polarization) is 
an experimental challenge. Recently, the idea was proposed to consider the nuclear bath at very 
low temperatures in the FM phase, which is mediated by many itinerant electons
via the RKKY-interaction. \cite{Reininghaus06,Simon07,Braunecker09,Smerat09}
In Ref.~\onlinecite{Braunecker09} it was shown that the Coulomb interaction in a 2d electron gas leads to an 
increased critical temperature of order $T\sim 1mK$ for the nuclear spins, which might be feasible 
in experiments. In Ref.~\onlinecite{Simon07} a $C^{13}$ carbon nanotube was
studied. By approximating the conduction electrons by a Luttinger liquid and treating the large effective 
nuclear spins classically, the transition temperature between a helically 
ordered (FM for finite systems) and unordered spin lattice was calculated. \cite{Simon07}
It could be shown that a finite long-ranged Coulomb interaction is required to have a 
finite transition temperature, \cite{Braunecker09} which is consistent with the Mermin-Wagner 
Theorem \cite{Wagner66} and its recent extension.\cite{Bruno01} Taking backaction effects of 
the nuclear lattice on the electron spins into account increases the transition temperature 
by another order of magnitude. This makes the KLM interesting for experiments, which are 
always performed at finite temperature.

These developments motivate the study of the KLM in the presence of a finite
Coulomb interaction between the itinerant electrons. The simplest extension to the 
KLM in terms of lattice models is the onsite Coulomb interaction $U$. In the case of 
half-filling a finite $U$ leads to the opening of a spin and charge gap. \cite{Shibata96} 
This work has been extended within a continuum Luttinger liquid approach to arbitrary 
fillings solved by bosonization. \cite{Gulacsi05} Lattice effects have been accounted 
for by means of a phononic field and therefore there is no real lattice involved in 
those calculations. Still, the authors of Ref.~\onlinecite{Gulacsi05} find the 
interesting result of a shift of the phase boundary between FM and PM phase, as expected.

In this paper, we use the density matrix renormalization group method 
\cite{White1992, White1993, Schollwoeck2005, Schollwoeck2010} (DMRG) to study ground state and dynamical 
properties of the one dimensional KLM for local spins with spin $1/2$ including onsite and nearest 
neighbor Coulomb interaction. Our method benefits from being numerically exact, 
acting in the lattice space without any approximations and taking all backaction 
effects of the local spin lattice on the conduction electrons automatically into 
account. Furthermore it allows for calculations in a broad parameter regime and works 
especially well for one dimensional systems with open boundary conditions and finite 
lattices. Here we are particularly interested in finite lattices, since nanoscale 
systems have finite sizes and show corresponding effects. 

From ground state calculations we show that onsite Coulomb interaction lowers the 
value of $J$ required for a transition from a PM to a FM ground 
state. For small $n \lesssim 0.4$ nearest neighbor Coulomb interaction $V$ acts the 
same way on the magnetic order as $U$ does. For $n \gtrsim 0.4$ they compete with each 
other. As a different sensor of magnetic order we utilize the static electron spin 
susceptibility. For the PM phase a peak at $2k_F$ is expected (which diverges for 
$L \rightarrow \infty$), while for the FM order a minimum at the smallest possible 
quasimomentum $q$, which is finite for finite lattices, should emerge. This was 
stated similarly in Refs. \onlinecite{Braunecker09},\onlinecite{Simon07} for small 
coupling constants $J$.

Finally, we calculate the quasiparticle life-time broadening $\Gamma_{+}$ 
of an electron, its spin oriented in the opposite direction than that of all other 
electrons in the ground state. In Ref.~\onlinecite{Smerat09} it was shown in the 
FM phase and for electronic densities below half-filling that the effective 
interaction between spinpolaron states is weak proving that spinpolaron (spin-down) 
states are indeed well-defined quasiparticles with small life-time broadening 
$\Gamma_{-}$ even in the presence of many electrons. However we will 
show here that the spin relaxation and decoherence rates will be dominated by the 
life-time broadening $\Gamma_{+}$ of the opposite spin-up state, which is higher in energy. 
We will consider a single spin-up electron with quasimomentum $k$ on top of the FM ground 
state of the 1d KLM. Although this spin has the same direction as the underlying local spins and, 
thus, can not decay by direct exchange with the local spins, we find that $\Gamma_{+}$ is dominated by the
effective exchange interaction with the sea of spinpolaron spin-down states in the system. 
As a consequence, $\Gamma_{+}$ turns out to be much larger than $\Gamma_{-}$ and dominates
the spin relaxation as well as the spin decoherence rate (the pure dephasing term arising
from the life-time broadening $\Gamma_{-}$ of the spin down spinpolaron state is negligible). 
We analyze the life-time broadening $\Gamma_{+}$ depending on $J$, $U$, $n$ and the 
quasimomentum $k$ and give explanations for the observations. Although the spin relaxation rate
increases significantly in the presence of many electrons we will show in appropriate parameter regimes
that the spin relaxation rate can be several order of magnitudes smaller in the FM phase
compared to the PM phase.

%=====================================================
%
%                   MODEL
%
%=====================================================
\section{Model}
The Hamiltonian of the KLM with Coulomb interaction is sketched in Fig.~\ref{fig:klm} and defined as
\begin{equation}
\label{eq:KLM}
\begin{split}
H = & -t\sum_{\sigma,i=1}^{L-1} \left( c_{i\sigma}^{\dagger} c_{i+1\sigma} + 
c_{i+1\sigma}^{\dagger} c_{i\sigma} \right) + J \sum_{i=1}^L \bvec{S}_i \cdot \bvec{s}_i \\
 &+ U \sum_{i=1}^L n_{i \uparrow} n_{i\downarrow} + V \sum_{i=1}^{L-1} n_{i} n_{i+1}
\end{split}
\end{equation}
where $t$ is the hopping integral, $L$ the lattice size, $c_{i \sigma}^{(\dagger)}$ the 
electron annihilation (creation) operator at site $i$ with spin $\sigma$, $J>0$ the 
antiferromagnetic Kondo exchange coupling, $\bvec{S}_i$ the spin operator of the local spin at 
site $i$, $\bvec{s}_i$ the spin operator of the conduction electron at site $i$, $U$ 
the onsite Coulomb interaction constant, $n_{i \sigma} = c^{\dagger}_{i \sigma} c_{i \sigma}$, 
$V$ the nearest neighbor Coulomb interaction constant and $n_{i}= n_{i \uparrow} + n_{i \downarrow}$. 
All spins are considered to be spin $1/2$. We define the filling by $n=N/L$, where $N$
denotes the total number of itinerant electrons ($n=1$ corresponds to half-filling).
%=====================================================
%
%                   METHOD
%
%=====================================================
\section{Method}
\subsection{DMRG}
The DMRG method is a well established numerically exact method for the calculation 
of ground states, dynamical properties and time evolution of one dimensional lattice 
systems. Our algorithm is formulated in a matrix-product language \cite{McCulloch2007} 
and makes use of Abelian, e.g. particle number conservation ($U(1)$) and non-Abelian, 
e.g., total spin conservation ($SU(2)$), symmetries. Depending on the symmetry sector, 
the use of $SU(2)$ symmetries in addition to $U(1)$ symmetries allows for computations 
up to 10 times faster.

\subsection{Ground states}
Calculating the ground state of a given system is synonymous to finding the symmetry 
sector with its corresponding quantum numbers, where the energy is minimal. The ground 
state phase diagram of the KLM is shown in Fig.~\ref{fig:pd} in dependence of the Kondo 
constant $J$ and the filling $n$. Fixing $J$ and n leaves the total spin quantum number 
$S$ as the only free parameter, which distinguishes the order of the ground state, i.~e., 
$S=(L-N)/2$ complies with FM order of local spins and $S=0$ with PM 
order. We choose $SU(2)$ symmetry for the spin here, first for computational reasons and 
second it has the benefit that the states with different total spin quantum numbers are 
non-degenerate in this case, whereas in $U(1)$ symmetry a partial degeneracy in the total 
spin in the direction of quantization exists. Considering Coulomb interaction in addition, 
we have another two variables that have to be fixed in advance and this means we have a 
quadruple of variables $\{n, J, U, V\}$, or a four dimensional phase diagram.

\subsection{Susceptibility}
We calculate the static electron spin susceptibility $\chi(\omega=0)$
by means of Green's functions and the application of dynamical DMRG 
\cite{Kuhner1999, Jeckelmann2002} with \emph{GMRES}. \cite{Soos89, Ramasesha97} 
Details of our implementation can be found in Ref.~\onlinecite{Smerat09}.

The definition of the spin susceptibility is
\begin{equation}
\begin{split}
	\chi^{+-}_q(\omega)  =  
        -\frac{1}{L} \left[ \bra{0} \tilde{s}_q^+ 
          \frac{1}{H-E_0+\omega - i\eta} \tilde{s}_q^- \ket{0} + \right.\\ 
	\left. \bra{0} \tilde{s}_q^- \frac{1}{H-E_0-\omega + i\eta} \tilde{s}_q^+ \ket{0} \right]
\end{split}
\end{equation}
with (for open boundary conditions)
\begin{eqnarray*}
	\tilde{s}_{q} = \sum_{l=1}^{L} s_l \,\sin\left(\frac{q l \pi}{L+1}\right)\quad, 
\end{eqnarray*}
where $H$ is the Hamiltonian given in Eq. (\ref{eq:KLM}), $\ket{0}$ is the ground state of the system, 
and $E_0$ the ground state energy. $\eta$ is a finite artificial broadening factor, needed to 
avoid finite size effects \cite{Schollwoeck2005} and which can be choosen smaller with larger lattice size.

\subsection{Quasiparticle life-times}
\begin{figure}
	\includegraphics[width=8.6cm]{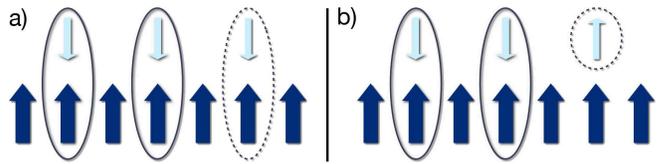}	
	\caption{\emph{(Color online)} (a) Sketch of a configuration with three spinpolarons, 
          each consisting of a delocalized spin singlet state with the local spins. 
	(b) Sketch of a configuration with two spinpolarons and one spin up electron. }
	\label{fig:qubit}
\end{figure}
In Ref.~\onlinecite{Smerat09} the quasiparticle life-time of the spinpolaron was 
calculated (cf. Fig.~\ref{fig:qubit}a), by evaluating the electronic Green's function 
in momentum and frequency space
\begin{multline}	
G_{k \sigma}(\omega+i\eta) = \\
	\frac{1}{\omega+i\eta-\left( \epsilon_0(k) - \mu + \Sigma_{\sigma}(k,\omega + i \eta) \right)},
\end{multline}
where $\omega$ is the energy, $\epsilon_0(k)$ the free electron dispersion relation, 
$\mu$ the electrochemical potential (which does not play a role in the calculation of 
broadenings of spectral densities) and $\Sigma_{\sigma}(k,\omega + i \eta)$ the complex 
self-energy. From the imaginary part of the self-energy, which is given by the broadening 
of the Lorentzian shaped peak in the spectral density 
$A_\sigma(k,\omega) = -(1/\pi)\: \text{Im} \: G_{k \sigma}(\omega) $ we can determine the 
quasiparticle life-time in dependence of all parameters. On the technical side, we use 
again the above mentioned GMRES method and calculate spectral densities as described 
in Ref.~\onlinecite{Smerat09}.

Basically, there exist four different scenarios for which the electronic quasiparticle 
life-time broadenings can be calculated assuming that in the FM ground state the local 
spins point up and the conduction electron spins point down (for large $J$ the most dominant
part of a spinpolaron state consists of a conduction electron pointing down with a small
admixture of the spin up state plus a local magnon): 
\begin{itemize}
\item \textbf{1} In the FM phase for a spin down electron (cf. Fig.~\ref{fig:qubit}a);
\item \textbf{2} In the FM phase for a spin up electron (cf. Fig.~\ref{fig:qubit}b); 
\item \textbf{3} and \textbf{4} are the corresponding cases for the PM phase. 
\end{itemize}
\textbf{1} corresponds to the spinpolaron life-time broadening $\Gamma_{-}$
and \textbf{2} to its natural counter part $\Gamma_{+}$.
\textbf{3} and \textbf{4} are identical, since the spins in the PM ground state have 
no specific direction. 

In addition to Ref.~\onlinecite{Smerat09} we calculate here the life-time broadening $\Gamma_{+}$.
As shown in this paper this rate is very large in the presence of many electrons, 
$\Gamma_{+}\gg \Gamma_{-}$, and, as a consequence, dominates the spin
relaxation and decoherence rates, as can be understood from the following qualitative analysis.
The two many-body spin states $|\pm\rangle$ depicted in Fig.~\ref{fig:qubit} are not exact eigenstates
but are expected to be part of a sharp many-body continuum with long
life-times. The spin down state $|-\rangle$ is protected from magnon
absorption and emission processes since the spinpolarons
can lower their energy by the entanglement with
the local spins in a singlet state. Only virtual processes and 
weak spinpolaron-spinpolaron interactions lead
to a small broadening $\Gamma_{-}$ of this state, as shown in detail
in Ref.~\onlinecite{Smerat09}. The spin-up state $|+\rangle$ is
protected due to the spin polarization of the local spins.
Due to effective exchange interaction between the spinpolarons 
and the spin-up electron mediated by the magnons, 
as discussed in detail in this paper in section \ref{sec:qplifetime},
this state has a life-time broadening $\Gamma_{+}\gg\Gamma_{-}$.
Denoting the quasienergies of the two spin states by $E_{\pm}$, we get a
decay according to $\langle\pm|e^{-iHt}|\pm\rangle\sim e^{-iE_\pm t}e^{-(\Gamma_\pm/2)t}$.
To define the spin relaxation and decoherence rates, we 
introduce pseudo-spin operators 
$P_z=(1/2)(|+\rangle\langle +|-|-\rangle\langle -|)$
and $P_\pm=|\pm\rangle\langle\mp|$. Using spin conservation, we
obtain after a straigthforward calculation that
$\langle P_z(t)\rangle=(1/2)|\langle+|e^{-iHt}|+\rangle|^2$,
if the system is prepared at $t=0$ in the state $|+\rangle$,
and $\langle P_+(t)\rangle=(1/2)\langle+|e^{-iHt}|+\rangle^*
\langle-|e^{-iHt}|-\rangle$, if the system is prepared in
the state $(1/\sqrt{2})(|-\rangle + |+\rangle)$ intially. As a result
we find for the two different initial preparations that
$\langle P_z(t)\rangle \sim e^{-\Gamma_1 t}$ and
$\langle P_+(t)\rangle \sim e^{i\Delta t}e^{-\Gamma_2 t}$,
where $\Delta=E_+-E_-$ is the quasienergy splitting and
the spin relaxation/decoherence rates are given by
\begin{equation}
\label{eq:relaxation_decoherence_rates}
\Gamma_1=\Gamma_+ \quad,\quad 
\Gamma_2={1\over 2}\Gamma_1 + {1\over 2}\Gamma_- \quad.
\end{equation}
This result shows that the dominant part to $\Gamma_{1/2}$ is
given by the broadening $\Gamma_+$ of the spin-up state $|+\rangle$,
whereas the broadening $\Gamma_-$ of the spinpolaron state $|-\rangle$
enters only into the pure dephasing term of longitudinal fluctuations
and can be neglected.

\subsection{Dispersion relation}
The dispersion relation $\omega_\sigma(k)$ can be constructed from the resonance
of the single particle spectral density $A_\sigma(k,\omega)$ at $\omega=\omega_\sigma(k)$.
The number of $k$ values is restricted by the lattice size $L$.

%=====================================================
%
%                   RESULTS
%
%=====================================================
\section{Results}

In nearly all cases we have choosen $L=48$, which is suitable from two different points 
of view. First, physically, we are especially interested in finite systems, which would 
more closely resemble, e.g.,  nanotubes in the real world. And second, from the point of 
view of computational cost, it is not convenient to take larger systems into account, 
since we already needed up to 3000 DMRG states in some of the calculations, which is a 
large number considering the number of executed calculations. All calculations are done 
with high computational precision, partly up to machine precision. We set $t=1$ in all calculations.

\subsection{Phase diagram}
\label{sec:phase_diagram}
\begin{figure}[t]
	\includegraphics[width=8.6cm]{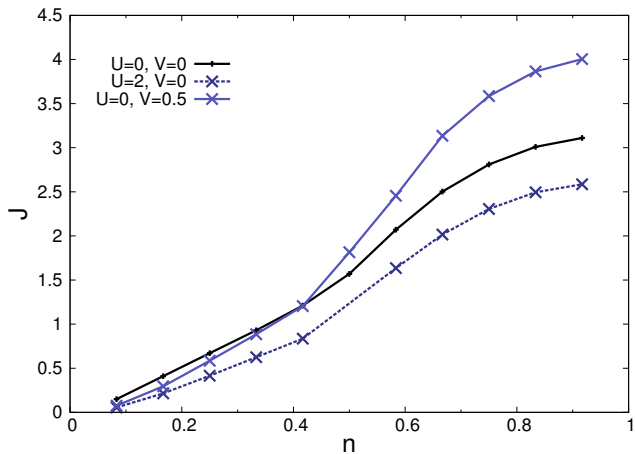}	
	\caption{\emph{(Color online)} Phase boundary between the FM (upper part) 
          and the PM (lower part) ground state of the Kondo lattice model with $L=48$ 
          for three different cases of Coulomb interaction.}
	\label{fig:pd}
\end{figure}
We will first investigate the influence of Coulomb interaction on the ground state of the 
Kondo lattice model. The phase diagram \cite{Tsunetsugu97}  of the KLM (without Coulomb interaction) 
is well established and shows two different phases, an FM and a PM one, 
see Fig.~\ref{fig:pd}. The PM phase lies in the lower-right triangular of the phase diagram and for 
all other values of $J$ and $n<1$ the KLM has an FM ground state. Especially for $N=1$ it was shown 
that the KLM is FM for any $J$.\cite{Sigrist91} As can be seen from Fig.~\ref{fig:pd}, applying a 
finite onsite Coulomb interaction shifts 
the phase boundary downwards for all values of $n$. This is consistent with the analysis of 
Ref.~\onlinecite{Simon07}, where a higher crossover temperature has been predicted in the presence
of Coulomb interaction. However, we note that the two mechanism are quite different. Whereas in
Ref.~\onlinecite{Simon07} the local nuclear spins have been treated quasiclassically due to their
large effective spin, the present analysis is in the full quantummechanical regime of local spins
with spin $1/2$. Roughly speaking the present result is consistent with
the Stoner picture of ferromagnetism, where a finite Coulomb interaction leads
to the preference of a fully spin-polarized state for the itinerant electrons. This
state coincides with the qualitative picture of spinpolaron states pointing into the
opposite direction of the local spins, see Fig.~\ref{fig:qubit}a.

For finite nearest neighbor Coulomb interaction $V$ we find the qualitatively different result, 
that the phase boundary is shifted downwards for $n \lesssim 0.4$ and upwards for $n \gtrsim 0.4$ 
and therefore crosses the phase boundary of the KLM without Coulomb interaction. For small fillings 
this can be explained in the same way as for the onsite Coulomb interaction case. For filling $n>0.4$ 
the electrons are relatively close to each other and therefore strongly influenced by $V$. 
The possibility to occupy the same site with two electrons of opposite spin does not lead 
to an increasing energy due to Coulomb interaction and increases the kinetic energy at the same
time. Therefore, in this regime, the unordered state becomes more favorable.

Summarizing, the onsite and nearest neighbor Coulomb interaction are concurring for small $n<0.4$ 
and behave competitively for large $n>0.4$. These results are pictured in Fig.~\ref{fig:pd}: The 
solid blue line is the phase boundary of the non-interacting KLM. If Coulomb interaction is switched 
on, the phase boundary is lowered for all values of $n$ (dashed dark blue line). For $U=0$ and $V$ 
finite, the phase boundary is lowered for small $n$ and raised above the non-interacting case phase 
boundary for larger $n$.
\input{table01}
\begin{figure}[t]
	\includegraphics[width=8.6cm]{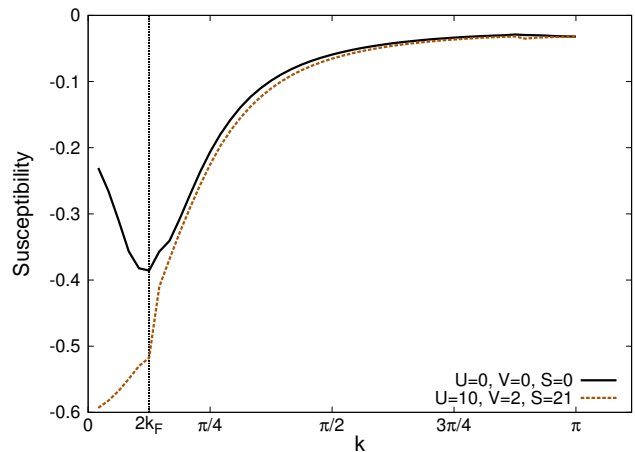}
	\caption{\emph{(Color online)} Static electron spin susceptibility $\chi(\omega=0)$ 
          for a Kondo lattice model with $L=48$, $N=6$ and $J=0.15$ with (dashed line) and without 
          (solid line) Coulomb interaction. The thin vertical line marks $2k_F$ in the PM phase.}
	\label{fig:susc}
\end{figure}

\subsection{Susceptibilities}
For small $J$ the order of the local spins manifests itself also in the static electron 
spin susceptibility. As was shown in Ref.~\onlinecite{Simon07} the effective coupling between 
the local spins for small $J$ is
\begin{equation}
	J_{RKKY} \propto -\chi^{\pm}(\omega=0, k, J, U).
\end{equation}
Therefore the order of the local spin lattice should correspond to the absolute maximum of the 
static electron spin susceptibility. In Fig.~\ref{fig:susc} we show this for two extreme cases 
with $L=48$ and $N=6$. The first case (solid black line in the figure) with $U=0$, $V=0$ 
has a PM ground state and shows the susceptibility in the non-interacting case. It 
has an absolute maximum at $k = 2k_F$. This evidences that for the chosen set of parameters 
the state indeed orders paramagnetically in a RKKY like fashion. If Coulomb interaction is 
switched on with $U=10$, $V=2$ (dashed brown line in the figure) the absolute maximum 
is at $k=0$. In this case FM order becomes dominant.

\begin{figure}[t]
	\includegraphics[width=8.6cm]{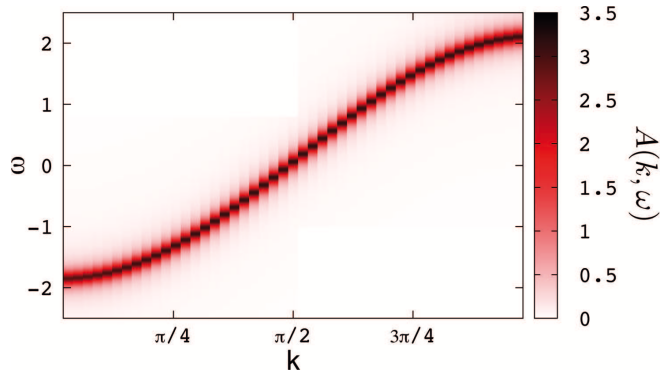}
	\caption{\emph{(Color online)} Dispersion relation of a $\uparrow$-electron 
          in a KLM with $J=0.5$, $N=4$ and $U=V=0$.}
	\label{fig:disp}
\end{figure}

\subsection{Dispersion relation}
We calculated the dispersion relation of a $\uparrow$-electron in a KLM with $L=48$, $N=4$, $J=0.5$ 
and $U=V=0$. The result is shown in Fig.~\ref{fig:disp}. It shows a cosine shaped dispersion, 
which leads to the conclusion that the electron behaves more or less like a free electron, only 
slightly affected by the presence of the local spin lattice. This can be explained by the fact
that a $\uparrow$-electron can not flip its spin directly by an exchange process with a local
spin due to spin conservation. In contrast, a $\downarrow$-electron can do so, leading to the
formation of spinpolarons, which can lower their energy by this process and obtain a larger
effective mass leading to a sharper dispersion relation. However, as shown in the next section,
the life-time broadening of $\uparrow$-electrons is generically larger than those of 
$\downarrow$-electrons, since the decay processes for spinpolarons start in higher order in $J$
than those for $\uparrow$-electrons.
\label{sec:qplifetime}
\begin{figure}[t]
	\includegraphics[width=8.6cm]{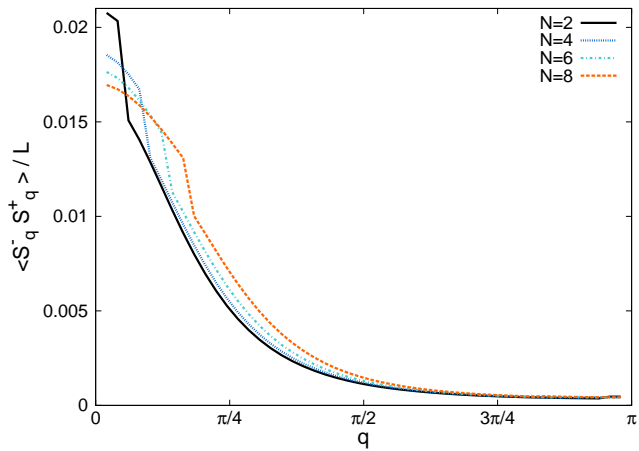}
	\caption{\emph{(Color online)} Magnon density in the KLM with $L=48$, $t=J=1$, $U=V=0$ 
          in dependence of the quasimomentum. The number of electrons is varied between 
          $2$ and $8$ in steps of $2$.}
	\label{fig:magndens}
\end{figure}
\begin{figure}[t]
	\includegraphics[width=8cm]{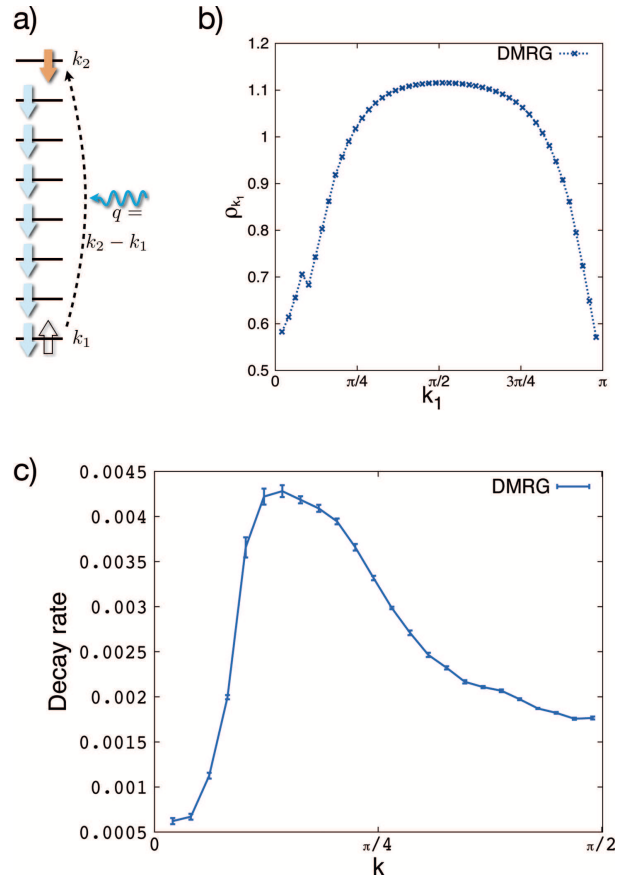}
	\caption{\emph{(Color online)} \emph{(a)} Simplified itinerant electron band structure in 
          $k$-space. Light blue electrons on the left side are electrons initially in the ground state 
          and electrons on the right side are additionally added to the ground state. The process shown 
          correspond to a spin flip of the added electron at $k=k_1$. After the spin flip, the electron 
          has opposite spin with $k=k_2$ and has absorbed a magnon with $q=k_2-k_1$.  \emph{(b)} 
          Accumulated magnon density $\rho_{k_1}$ as given in Eq. (\ref{eq:acc_magn_dens}), for 
          $L=48$, $N=4$ and $J=0.5$. \emph{(c)} Decay rates for $L=48$, $N=4$ and $J=0.5$ in dependence of $k$.}
	\label{fig:accmagn}
\end{figure}
\input{table02}

\subsection{Quasiparticle life-times}
From the electronic spectral density $A_\uparrow(k,\omega)$ we obtain the quasiparticle 
life-time broadenings $\Gamma_+$ in dependence of $J$, $U$, $k$ and $N$. As we 
calculate the Green's function $G_\uparrow(k,\omega)$ in frequency space, we obtain 
two branches: The $c_{k \uparrow}^{\dagger}$-- and the $c_{k \uparrow}$--branch, respectively. 
The first one corresponds to an additional electron placed in a certain $k$-mode and interacting 
with the other electrons and the local spins. The second type addresses the spin up-part of the 
already existing electrons in the system. Therefore the two branches address two different sets 
of states in the spectrum of the Hamiltonian. Here we are interested in the first case only, 
since we would like to know, what happens to a spin up electron brought into the system in addition 
to the other electrons.

\subsubsection{Decay rate dependence on $k$}
\label{sec:rate_k}
In Tab. \ref{lifetimes_kdep} we show decay rates of a spin up electron added to the $N=2$ ground state. 
For all sets of $U$ and $J$ we find that the decay rate increases with increasing $k$ as 
long as $k$ is smaller than $2k_F$. Here we give an explanation considering momentum conservation and
phase space arguments. In the FM ground state the lowest electronic orbitals in $k$ 
space are occupied up to $2k_F$ by the available electrons all with spin down. 
A state with wave vector $k$ has quasimomentum $\pm k$ due to the open boundary conditions.
An additionally superimposed spin up electron 
with a certain wave vector $k_1$ has to change to the state $k_2 > 2k_F$ in order to flip 
its spin, see Fig.~\ref{fig:accmagn}a. This decay channel can only happen if a magnon
is absorbed with wave vector $q=|k_1\pm k_2|$. Such magnons are present in the ground state
because each spinpolaron state consisting of a spin down electron with wave vector $k$ has a 
small admixture of spin up 
states with wave vector $|k\pm q|$ and a local magnon in state $q$. Smaller values of $k_1$ 
decreases the number of magnons with small wave vector $q=|k_1-k_2|$ to enable this process. 
This can be quantified by the magnon density per electron $m_q  = \left< S_q^- S_q^+ \right>/N$ 
(see Fig.~\ref{fig:magndens}) and further by the accumulated magnon density 
\begin{equation}
\label{eq:acc_magn_dens}
\rho_{k_1} = \sum_{\sigma=\pm}
\sum_{\substack{q=|k_1+\sigma k_2|\\ 0<q<\pi \,,\, 2k_F<k_2<\pi}} m_q \quad,
\end{equation}
which is shown in Fig.~\ref{fig:accmagn}b and clearly states that the number of suitable magnons 
increases with increasing $k_1$ even above $2k_F$ until it falls off finally. This result 
qualitatively reflects the decay rate for the spin up electron shown in Fig.~\ref{fig:accmagn}c
for a KLM with $L=48$, $N=4$, $J=0.5$ and $U=V=0$. The decay rate first increases for small $k$ as 
indicated by the accumulated magnon density. For values above $2k_F$ the decay rate even surpasses 
the values at $2k_F$ until it decreases finally for larger values of $k$. We note that this is only a
qualitative explanation since other decay channels involving absorption of many magnons are
present as well.

The discussed process for the decay of the spin up electron is essentially an exchange process
between a spin up electron in state $k_1$ and a spinpolaron in state $k$. The spinpolaron
provides the magnon with wave vector $q=|k_1\pm k_2|$ to flip the spin up electron from state 
$k_1\rightarrow k_2$, leaving the spinpolaron as a spin up electron in state $|k\pm q|$. As a
result, by mediation of a local magnon, the spins of two electrons have been exchanged, whereas 
the local spin lattice is unaffected. This spin exchange process is the essential process 
leading to a large life-time broadening of the spin-up electrons if many electrons are present
in the system. In contrast, the spinpolaron states have life-time broadenings, which are several
orders of magnitude smaller compared to those of the spin-up states. The reason is that the
spinpolaron-spinpolaron interaction is rather weak and can only be mediated via multi-magnon processes.

\subsubsection{Decay rate dependencies on $U, J, N$}
In this section we will explain how the quasiparticle decay rate of the spin up state
depends on $U,J$ and $N$ and why the found tendencies are to be expected. The results for these cases 
are shown in Tab. \ref{lifetimes}. 

Let us first consider the $J$-dependency. Picking one of the columns and considering only one of the 
two $U$-values we immediately recognize that the decay rate shrinks with decreasing $J$. The exchange 
strength $J$ determines the time scale on which spins will flip, therefore with decreasing $J$ 
flipping will be suppressed and the rate decreases. We note that this is different for the decay rate
of the spinpolaron, where an increasing $J$ stabilizes each polaron and makes it 
insensitive to interactions with other electrons. For small $J$ close to or even in the 
PM phase the decay rate of the spin up state increases notably, see $N=4$. This is 
natural, since in a paramagnetically ordered system many additional decay channels will open up.

Considering the $U$ dependence we find that with increasing $U$ the rate increases in most cases. 
In section \ref{sec:phase_diagram} we have found that an onsite Coulomb interaction has the 
tendency to order the local spins ferromagnetically. The additional spin up electron tries to 
align parallel to the other electrons to minimize interaction energy from the Coulomb potential. 
This infers a larger decay rate, if $U$ becomes larger. Therefore this tendency here complies 
with the influence of the onsite Coulomb interaction found above. Only when a finite $U$ triggers
the crossover from the PM to the FM phase, the rate decreases with 
increasing $U$, see $N=4$ and $J=0.3$. This is obvious since in the PM phase the phase
space arguments presented in section \ref{sec:rate_k} are no longer valid and many more
decay channels are possible.

If we increase the number of electrons $N$ in the system and keep the quasimomentum $k$ fixed we 
find that the rates decrease with increasing $N$, for small $N$ deep in the FM phase. 
This can be explained analog to the discussion in section \ref{sec:rate_k}. In the ground state, all 
initially available electrons fill the spinpolaron-band successively up to $2k_F$ mainly 
in the spin down state. An additional spin up electron can be added to any $k$-mode. In 
Tab.~\ref{lifetimes} we considered the lowest state $k=\pi/(L+1)$ in all cases. Considering one of the rows 
the electron number is increased from left to right and with each electron more in the ground 
state the respectively next higher $k$-mode is occupied by this additional electron. 
As a consequence, as shown in section \ref{sec:rate_k}, by increasing $N$ we decrease the number of magnons 
suitable for scattering processes and therefore the decay rate has to decrease. 
However, in competition to this effect, increasing $N$ means also approaching the PM 
phase. Then we expect that different and also more decay channels open up, which should lead to an 
increasing decay rate. This can be seen in Tab. \ref{lifetimes} for $J=0.8$ between $N=6$ 
and $N=12$. We have also calculated lifetimes for $N=7,9,10,11$ (not shown), showing that
the decay rates are monotonically increasing with increasing $N$ for large $N$. For values of $N$ close 
to half-filling of the conduction band and large values of $J$, such that we can switch 
between PM and FM phase, we find decay rates of the order of $0.01$.
As a consequence, the decay rate depends nonmonotonically on $N$, it decreases for small
values of $N$ deep in the FM phase and increases for larger values of $N$ when the
PM phase is approached.  

Nonetheless we find the \emph{sweet spot} of the system by \emph{decreasing} the number of electrons 
going from $N=4$ to $N=3$ electrons at $J=0.1$. There we find that the decay rate of the 
spin up electron decreases by two orders of magnitude when comparing the rates in the
PM and FM phase. Still it is important to note that 
a minimum number of electrons in the system is important to maintain the FM 
order, especially at finite temperatures.

%=====================================================
%
%                  Summary
%
%=====================================================
\section{Discussion}
In this work we discussed the phase diagram and the spin relaxation
properties of the 1d spin-$1/2$ Kondo lattice model with Coulomb 
interaction. We found that a finite onsite or nearest neighbor
interaction favors a FM order of the local spin lattice
for small enough electronic densities.
This gives further strong support to the analysis of 
Refs.~\onlinecite{Braunecker09},\onlinecite{Simon07}, where similiar results have
been found in 2d semiconductor systems and $C^{13}$ carbon nanotubes.
It provides a pathway to achieve a spontaneous and full polarization 
of the nuclear spins by lowering the temperature below the critical
one. This configuration is desirable for applications in quantum
information processing, since it reduces the spin relaxation and
decoherence rates of the electronic spins. It is important to notice
that a finite crossover temperature can only be expected, if the
density of electrons is finite. Thus, many electrons are 
necessary to achieve the FM state. Once the FM
state is achieved, one can in principle perform quantum information
processing by realizing quantum dots with external gates on time
scales which are small compared to the time the nuclear spins need
to return to the PM phase. If this is possible one
can effectively realize a system consisting of one single electron
$N=1$ in contact with a ferromagnetically ordered nuclear spin
lattice. In this case the spin up state and the spinpolaron are
exact eigenstates, i.e., the ideal situation with $\Gamma_\pm=0$ is
achieved.
In this paper we discussed the spin relaxation properties for
$N>1$, i.e. we analysed the question whether the spins in a
many-body system could possibly be used as candidates for spin
qu-bits. In Ref.~\onlinecite{Smerat09} we already found that
spinpolarons are indeed very long living states, indicating that
the spinpolaron-spinpolaron interaction is rather weak. However,
in this paper we found that the spin up state is strongly influenced
by exchange interaction between the spin up and spinpolaron states.
This exchange process does not require any finite energy and,
therefore, can not even be suppressed by application of a finite
magnetic field. We analysed in detail the dependence of $\Gamma_+$
on the Coulomb interaction $U$, the exchange interaction $J$,
the particle number $N$ and the quasimomentum $k$. In the 
FM phase we found that the rate decreases for
smaller values of $U$, $J$, $k$, and larger values for $N$,
unless we approach the PM phase. For appropriate
parameter sets we have shown that the life-time of spin up states 
can be two orders of magnitude larger in the FM phase
than in the PM phase. However, compared to the 
life-time of spin down spinpolaron states, their 
life-time is orders of magnitudes smaller, regardless 
of the chosen parameter regime in the FM phase.

%
%=====================================================
%
%               Acknowledgments
%
%=====================================================
%
\begin{acknowledgments}
We thank D. Loss for valuable discussions. 
H. Schoeller, U. Schollw\"ock and S. Smerat acknowledge the support from the 
DFG-Forschergruppe 912 on \lq\lq Coherence and relaxation properties of electron spins''.
\end{acknowledgments}
%
%=====================================================
%
%                  Bibliography
%
%=====================================================
%

%
\end{document}

%% file: table01.tex
\begingroup
\squeezetable
\begin{table*}%[H] add [H] placement to break table across pages
\caption{\label{lifetimes_kdep}$k$-dependence of relaxation rates for $N=2$ for different values of $J$ and $U$. }
\begin{ruledtabular}
\begin{tabular}{l | llll}
k $\left[ \pi/(L+1) \right]$ & 1& 2 & 3 & 4 \\
\hline
J=0.5, U=0 & $0.00097 \pm 0.00003$ & $0.00128 \pm 0.00002$ & $0.00166 \pm 0.00003$ & $0.00204 \pm 0.00005$\\

 \hline
J=0.5, U=0.2 & $0.00220 \pm 0.00009$ & $0.00299 \pm 0.00005$ & $0.00403 \pm 0.00005$ & $0.0048 \pm 0.0001$ \\

 \hline
 J=0.3, U=0 & $0.00035 \pm 0.00001$ & $0.000470 \pm 0.000004$ & $0.00066 \pm 0.00002$ & $0.00077 \pm 0.00002$\\

 \hline
 J=0.3, U=0.2 & $0.00146 \pm 0.00006$ & $0.00198 \pm 0.00004$ & $0.00280 \pm 0.00007$ & $0.00331 \pm 0.00008$ 

\end{tabular}
\end{ruledtabular}
\end{table*}
\endgroup

%% file: table02.tex
\begingroup
\squeezetable
\begin{table*}[t]%[H] add [H] placement to break table across pages
\caption{\label{lifetimes}Relaxation rates in dependence of the electron number $N$, $J$ and $U$. $k$ is set to the lowest possible value $k= \pi/(L+1)$. The given number of $N$ is the number of electrons taken into account during the ground state calculations, i.e., the spin up electron is in addition to this number. (p) mark parameters, which correspond to the paramagnetic phase.}
\begin{ruledtabular}
\begin{tabular}{l | lllllll}
N & 1 & 2 & 3 & 4 & 6 & 12\\
\hline
J=1.0, U=0 &  & &  $0.00263 \pm 0.00012$ & $0.00199 \pm 0.00013$ & $0.00185 \pm 0.00017$ & $0.00085 \pm 0.00018$\\
\hline
J=1.0, U=0.2 &  &  &  $0.00432 \pm 0.00070$ & $0.00294 \pm 0.00019$ & $0.00249 \pm 0.00027$ & $0.0011 \pm 0.0002$\\
\hline
J=0.8, U=0 &  &  &  $0.00188 \pm 0.00008$ & $0.00130 \pm 0.00009$ & $0.00087 \pm 0.00009$ & $0.00213 \pm 0.00024$\\
\hline
J=0.8, U=0.2 &  &  &  $0.00303 \pm 0.00021$ & $0.00221 \pm 0.00017$ & $0.00184 \pm 0.00017$ & $0.00238 \pm 0.00038$\\
\hline
J=0.6, U=0 &  &  &  $0.00115 \pm 0.00005$ & $0.00078 \pm 0.00005$ & $0.00081 \pm 0.00007$ & $\text{(p)} 0.00508 \pm 0.00031$\\
\hline
J=0.6, U=0.2 &  &  &  $0.00239 \pm 0.00012$ & $0.00156  \pm 0.00012$ &  & \\
\hline
J=0.5, U=0 & $0.00104  \pm 0.00002$ & $0.00097 \pm 0.00003$ &  $0.00082 \pm 0.00003$ & $0.00062 \pm 0.00004$ & $0.00066 \pm 0.00005$ &\\
 \hline
J=0.5, U=0.2 & $0.00233 \pm 0.00004$  & $0.0022 \pm 0.0001$ & $0.00205 \pm 0.00008$ & $0.00126 \pm 0.00011$ & $0.00141  \pm 0.00013$ & \\
 \hline
J=0.5, U=0.4 &   &  &  &  & $0.00142 \pm 0.00021$ & \\
 \hline
J=0.5, U=0.6 &   &  &  &  & $0.00245 \pm 0.00018$ & \\
 \hline
J=0.5, U=0.8 &  & &  &  & $0.00384 \pm 0.00045$ & &\\
 \hline
 J=0.3, U=0 & $0.00041 \pm 0.00001$ & $0.00035 \pm 0.00001$ & $0.00033 \pm 0.00001$ & $\text{(p) } 0.00144 \pm 0.00010$ & $0.00150 \pm 0.00020$ & \\
 \hline
 J=0.3, U=0.2 & $0.00158 \pm 0.0003$ & $0.00146 \pm  0.000016$ & $0.00123 \pm 0.00008$ & $0.00084 \pm 0.00007$  & & \\
 \hline
 J=0.1, U=0.0 & $0.000030 \pm 0.000001$ &  & $0.00004 \pm 0.0000009$ & $\text{(p) }0.00593 \pm 0.00024$ & & \\
 \hline
 J=0.1, U=0.2 &   &  & $0.00061 \pm 0.00003$& $\text{(p) }0.00460 \pm 0.00011$  &  & 
\end{tabular}
\end{ruledtabular}
\end{table*}
\endgroup